\documentclass[aps,prb,twocolumn,floatfix,showpacs]{revtex4}
\usepackage{amsmath}
\usepackage{graphicx}

\renewcommand{\r}{{\bf r}}
\renewcommand{\k}{{\bf k}}
\newcommand{\rhat}{{\bf \hat{r}}}
\newcommand{\khat}{{\bf \hat{k}}}

\begin{document}

\title{Filtering a distribution simultaneously in real and Fourier space}

\author{ Eduardo Anglada and Jos\'e M. Soler }
\affiliation{ Departamento de F\'{\i}sica de la Materia Condensada, 
              C-III, Universidad Aut\'{o}noma de Madrid,
              E-28049 Madrid, Spain }

\date{\today}

\begin{abstract}
   We present a method to filter a distribution so that it is
confined within a sphere of given radius $r_c$ and, simultaneously, 
whose Fourier transform is optimally confined within a sphere of
radius $k_c$.
   Our procedure may have several applications in the field of 
electronic structure methods, like the generation of optimized 
pseudopotentials and localized pseudocore charge distributions.
   As an example, we describe a particular application within the 
SIESTA method for density functional calculations, in removing the 
spurious rippling of the energy surface generated by the 
integrations in a real space grid.
\end{abstract}



\maketitle

\section{Introducction}

   It is well known that the mean quadratic widths, in real and Fourier 
space, $\Delta r^2$ and $\Delta k^2$, of a distribution in a space of 
$n_D$ dimensions, obey the uncertainty relation 
$\Delta r \Delta k \ge n_D$.
   The equal sign applies to a spherically symmetric gaussian 
distribution, which is therefore optimally confined in phase space
in the least squares sense.
   In many practical cases, however, we are interested in distributions
that are strictly confined within a sphere of given radius (i.\ e.\
defined to be strictly zero outside that sphere) and, simultaneously,
optimally confined within another sphere in Fourier space.
   This occurs when, to be computationally efficient, we use
distributions defined only within a finite sphere and we must limit 
also their Fourier transforms to a finite number of plane waves.
   In order to calculate those Fourier components, we frequently
must perform a discrete Fourier transform using a finite number of grid
points, and we want to avoid as much as possible the resulting aliasing 
effects~\cite{NumericalRecipes}.
   Such a situation occurs, for example, in the efficient computation of
Ewald sums, and in the particle-mesh method~\cite{Hockney-Eastwood1988}.
   Within the field of electronic structure calculations, this problem
occurs in the real-space formulation~\cite{KingSmith-Payne-Lin1991} 
of Kleinman-Bylander pseudopotentials~\cite{Kleinman-Bylander1982},
and of pseudocore charge distributions of ultrasoft pseudopotentials
\cite{Vanderbilt1990}.

   In the specific case of the SIESTA density functional method
\cite{Ordejon-Artacho-Soler1996,Soler2002}, this problem arises in the
evaluation, using a real-space grid, of matrix elements involving 
strictly localized basis orbitals and neutral-atom potentials.
   Those integrals generate an artificial rippling of the total energy, 
as a function of the atomic positions relative to the grid points
(the so-called eggbox effect), which complicates considerably the 
relaxation of the geometry and the evaluation of phonon frequencies 
by finite differences.
   In other grid-based methods~\cite{Beck2000}, this problem is 
generally solved by filtering the atomic pseudopotentials
~\cite{Briggs-Sullivan-Bernholc1996}, typically by multiplying them 
by an ad-hoc filter function in Fourier space~\cite{NumericalRecipes}.
   Here we present a new method for optimal filtering and its 
application to solve the eggbox problem in SIESTA.

\section{Optimized filtering method}

   We will study only the specific case of three dimensions, but
the extension to one or two dimensions is obvious.
   Consider an initial distribution of the form
\begin{equation}
F(\r) = \left\{
  \begin{array}{ll} 
    F(r) Y_l^m(\rhat)  & \mbox{if $r \le r_c$} \\
    0                  & \mbox{otherwise}
  \end{array}
\right.
\label{Fofr}
\end{equation}
where we are using the same symbol $F$ for $F(\r)$ and its radial
part $F(r)$, since it does not lead to any confusion.
   $Y_l^m(\rhat)$ is a real spherical harmonic.
   The Fourier transform of $F(\r)$ is 
\begin{equation}
G(\k) \equiv \frac{i^l}{(2 \pi)^{3/2}} \int d\r^3 e^{-i \k \r} F(\r) 
      = G(k) Y_l^m(\khat)
\label{Gofk}
\end{equation}
where we have introduced the factor $i^l$ to make $G(\k)$ real, and
\begin{equation}
G(k) = \frac{4 \pi}{(2 \pi)^{3/2}}  
  \int_0^{r_c} dr ~r^2 j_l(kr) F(r)
\label{GofF}
\end{equation}
where $j_l(x)$ is a spherical Bessel function.

   In general $G(k)$ will be nonzero for any value of $k$.
   If we want to filter it out for $k>k_c$, the most 
straightforward procedure is to multiply it by a step function
and then to perform the inverse Fourier transform:
\begin{equation}
F(r) \leftarrow \frac{4 \pi}{(2 \pi)^{3/2}}  
  \int_0^{k_c} dk ~k^2 j_l(kr) G(k).
\label{FofG}
\end{equation}
   The new $F(r)$ will no longer be strictly zero for $r>r_c$
but we may suppress those components and iterate the procedure.
   As a result, only the most confined components, in real and
reciprocal space, will survive.

   To annalize more rigorously the decomposition of $F(r)$ into
more and less confined components, let us define $x \equiv r/r_c$,
$y \equiv k/k_c$, $f(x) \equiv x F(x r_c)$,  
$g(y) \equiv y G(y k_c)$, $\kappa \equiv k_c r_c$, and 
$K(x,y) \equiv \sqrt{2 \kappa / \pi} ~\kappa x y ~j_l(\kappa x y)$.
   Then, substituting in (\ref{GofF}) and (\ref{FofG}), 
one iteration of the filtering procedure is given by
\begin{equation}
f(x) \leftarrow \int_0^1 dx' K^2(x,x') f(x')
\label{foff}
\end{equation}
where
\begin{equation}
K^2(x,x') \equiv \int_0^1 dy K(x,y) K(y,x').
\label{K2ofK}
\end{equation}

   If $f(x)$ were already a perfectly confined function in both
real and reciprocal space, it would not be affected by the filtering
procedure (\ref{foff}), i.\ e.\ it would be an eigenfunction of the 
filtering kernel $K^2$ with eigenvalue one.
   In practice, the uncertainty principle forbids simultaneous perfect 
confinement in real and Fourier space, and the filtered $f(x)$ will
unavoidably `leak' somewhat outside $x>1$ and its norm within $x \le 1$ 
will no longer be one.
   In fact, if $\phi(x)$ is an eigenfunction of $K^2$, with norm
equal to one within $x \le 1$, its eigenvalue $\lambda^2$ gives directly
its norm after filtering, since the effect of filtering is just 
a multiplication by $\lambda^2$:
\begin{equation}
\phi(x) \leftarrow \int_0^1 dx' K^2(x,x') \phi(x') = \lambda^2 \phi(x)
\label{eigsystem2}
\end{equation}
   Thus, we may perform an efficient filtering, without the need of 
iteration, by expanding the original function in terms of the complete 
basis of eigenfunctions of $K^2$, keeping only those with eigenvalues
sufficiently close to one.
   Since, it is clear that the eigenfunctions $\phi_i(x)$ of $K(x,y)$, 
with eigenvalues $\lambda_i$, are also eigenfunctions of $K^2(x,x')$,
with eigenvalues $\lambda_i^2$, we may work with the simpler
eigenvalue problem
\begin{equation}
\int_0^1 dy K(x,y) \phi_i(y) = \lambda_i \phi_i(x)
\label{eigsystem}
\end{equation}
   Notice that, since $K(x,y)$ is the Fourier-transform kernel, the 
eigenfunctions $\phi_i(x)$ have the same shape in real and reciprocal
space.
   This is not true in general for the filtered function $f(x)$, 
which is a combination of eigenfunctions with eigenvalues $\lambda_i$
close to either +1 or -1, which either change sign or not when
Fourier transformed.

   In order to solve (\ref{eigsystem}), it is convenient to 
expand $K(x,y)$ and $\phi_i(x)$ in a basis of functions
in the interval $[0,1]$.
   The simplest basis is that of powers of $x$.
   From the Taylor expansion of $j_l(x)$ at $x=0$ we find
$K(x,y) \simeq \sum_{n=0}^N K_n x^{2n+l+1} y^{2n+l+1}$, where
\begin{equation}
K_n = \sqrt{\frac{2 \kappa}{\pi}} 
      \frac{(-1)^n \kappa^{2n+l+1}}{(2n)!! (2n+2l+1)!!}.
\label{Kn}
\end{equation}
   Then making $\phi_i(x) = \sum_{n=0}^N \phi_{i n} x^{2n+l+1}$, 
Eq.~(\ref{eigsystem}) becomes
\begin{equation}
\sum_{m=0}^N \frac{K_n}{2n+2m+2l+3} \phi_{i m} = \lambda_i \phi_{i n}.
\label{eigsys}
\end{equation}

\begin{figure}[htpb]
\includegraphics[width=0.9\columnwidth,clip]{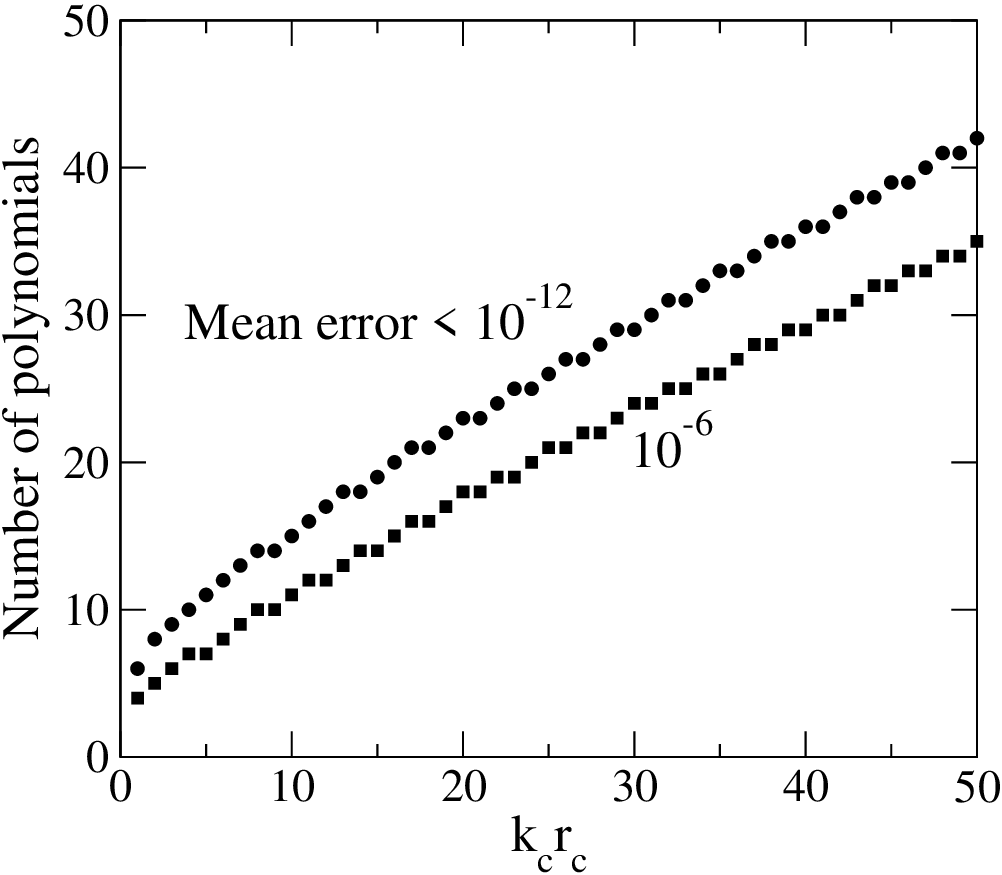}
\caption{
   Number of polynomials required to obtain a root mean square error 
of the expansion of $x j_0(x)$ in the interval $0<x<k_c r_c$.
   $j_0(x)$ is a spherical Bessel function with $l=0$.
}
\label{jlfit}
\end{figure}

   In practice, we have found numerically more accurate, stable,
and efficient (requiring a lower $N$) to expand $K(x,x')$ in 
orthonormal Legendre polynomials ~\cite{NumericalRecipes} $P_n(x)$ 
in the interval $0 \le x \le 1$.
   Taking into account the parity $l_p=\mod(l,2)$ of $j_l(x)$:
\begin{equation}
K(x,y) \simeq \sum_{n,m=1}^N K_{nm} P_{2n-l_p-1}(x) P_{2m-l_p-1}(y).
\label{Kexp}
\end{equation}
   The kernel coefficients $K_{nm}$ may be calculated by 
integration in a Gauss-Legendre~\cite{NumericalRecipes} set of 
points $x_\alpha$ and weights $w_\alpha$:
\begin{eqnarray}
K_{nm} &=& \int \int_0^1 dx ~dy K(x,y) P_{2n-l_p-1}(x) P_{2m-l_p-1}(y)
   \nonumber \\
&=& \sum_{\alpha,\beta=1}^{N-l_p} w_\alpha w_\beta K(x_\alpha,y_\beta) 
     P_{2n-l_p-1}(x_\alpha) P_{2m-l_p-1}(y_\beta)
\label{Knm}
\end{eqnarray}
   The required number $N$ of polynomials is determined by the
convergence of the expansion
$x j_l(x) \simeq \sum_{n=1}^N j_{ln} P_{2n-l_p-1}(x)$
in the interval $0 \le x \le \kappa$.
   Figure~\ref{jlfit} shows the number of polynomials $N$ required
to obtain a given error in the expansion, as a
function of $\kappa$, for $l=0$.
   The $l$-dependence of the error is very small and, 
as a rule of thumb, we use $N=\mbox{int}(10+0.65\kappa)$.

\begin{figure}[htpb]
\includegraphics[width=0.9\columnwidth,clip]{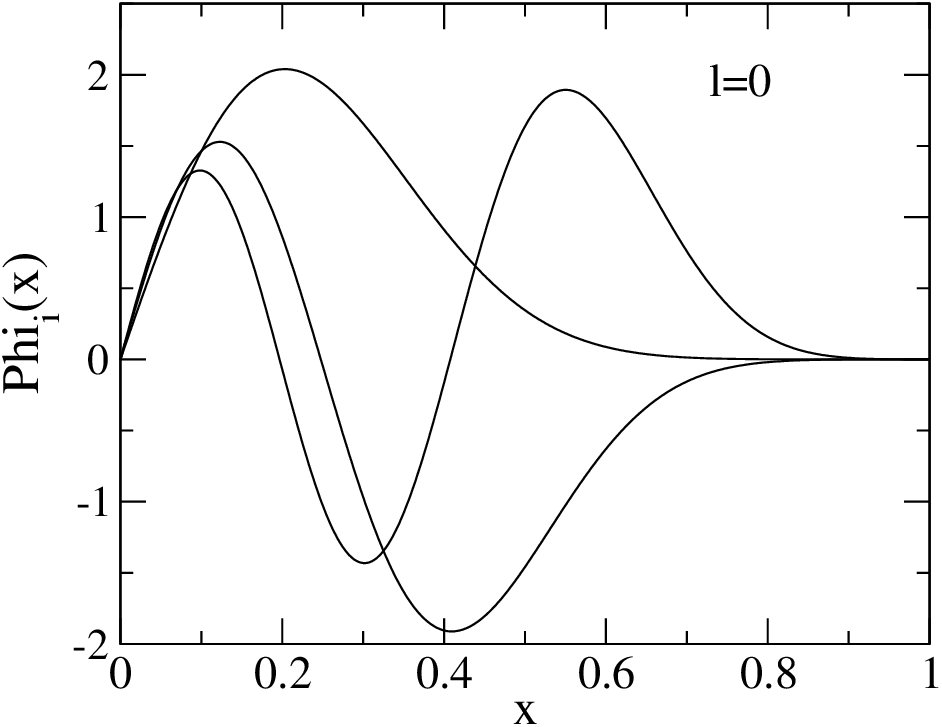}
\includegraphics[width=0.9\columnwidth,clip]{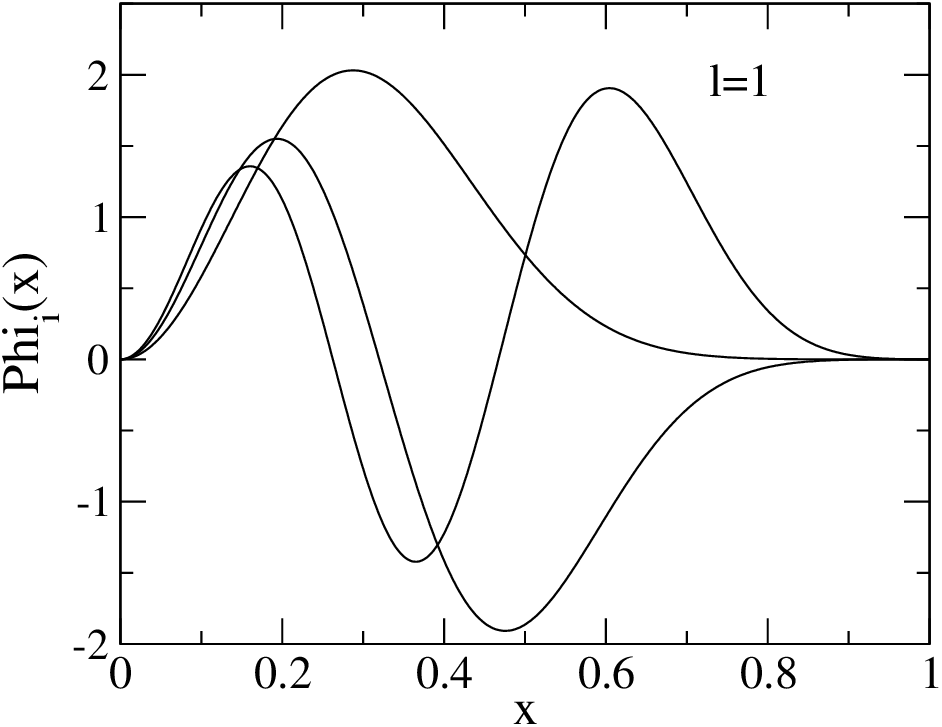}
\caption{
   First few eigenfunctions (with highest eigenvalues) of the filter
kernel $K^2(x,x')$ for $\kappa \equiv k_c r_c = 25$.
   Divided by $r$, they give the radial part of the distributions,
with angular momentum $l$,
that are most localized in a real-space sphere of radius $r_c$ and
simultaneously in a reciprocal-space sphere of radius $k_c$.
}
\label{phi}
\end{figure}

\begin{figure}[htpb]
\includegraphics[width=0.9\columnwidth,clip]{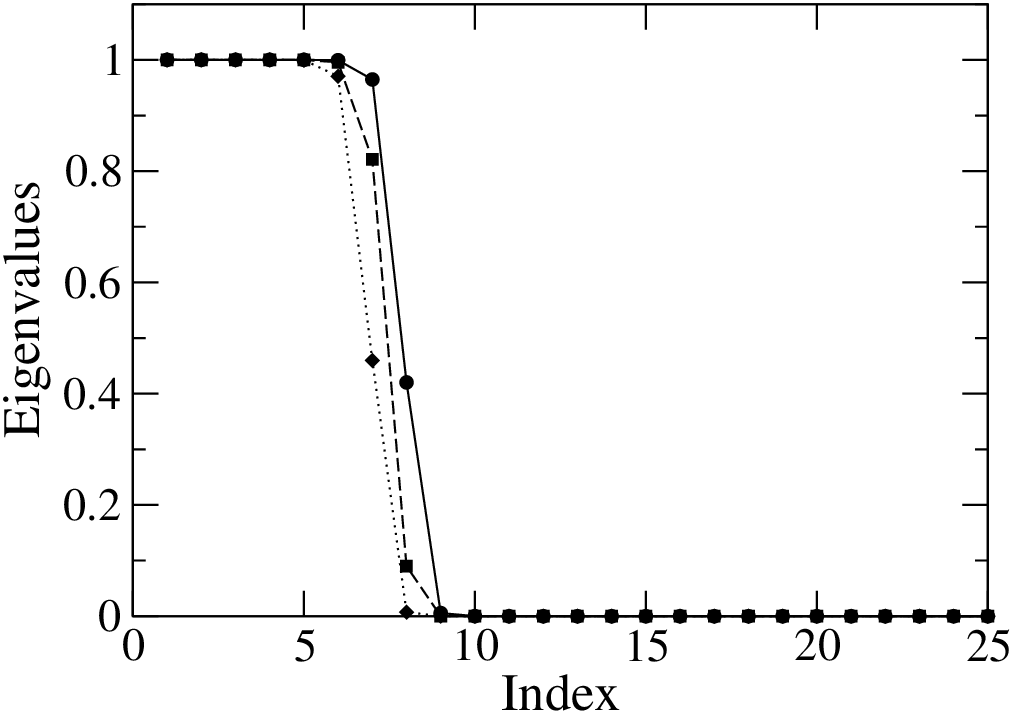}
\caption{
   Eigenvalues of the filter kernel $K^2(x,x')$ for $\kappa=25$
and $l=0$ (circles and full line), $l=1$ (squares and dashed line),
and $l=2$ (diamonds and dotted line).
}
\label{eigval}
\end{figure}

   Figure~\ref{phi} plots the first eigenfunctions $\phi_i(x)$ of
the filter kernel $K^2(x,y)$ for a typical value of $\kappa$, 
and figure~\ref{eigval} shows all the eigenvalues $\lambda_i^2$
up to $N$.
   It may be seen that there is a rapid transition between the
eigenvalues which are very close to 1 and those close to 0.
   It is then straightforward to select the $M$ eigenfunctions whose
eigenvalues are above some threshold, say $\lambda_i^2 > 0.99$, for
the expansion of the filtered function:
\begin{equation}
f(x) \leftarrow \sum_{i=1}^M f_i \phi_i(x)
\label{fexp}
\end{equation}
\begin{equation}
f_i = \sum_{\alpha=1}^{N-l_p} w_\alpha \phi_i(x_\alpha) f(x_\alpha).
\label{fi}
\end{equation}

\begin{figure}[htpb]
\includegraphics[width=0.9\columnwidth,clip]{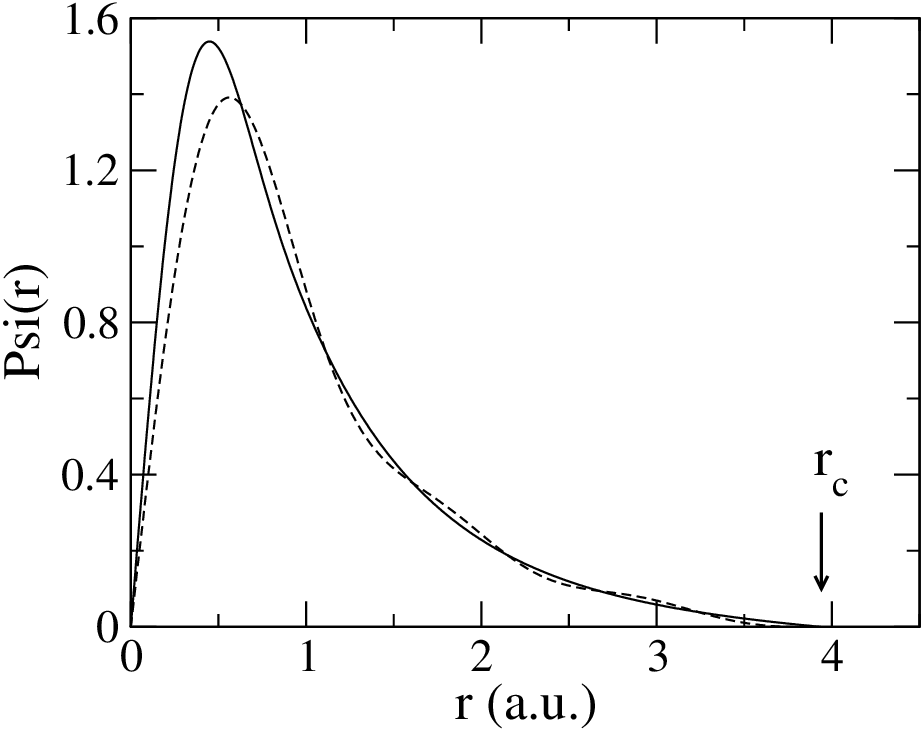}
\includegraphics[width=0.9\columnwidth,clip]{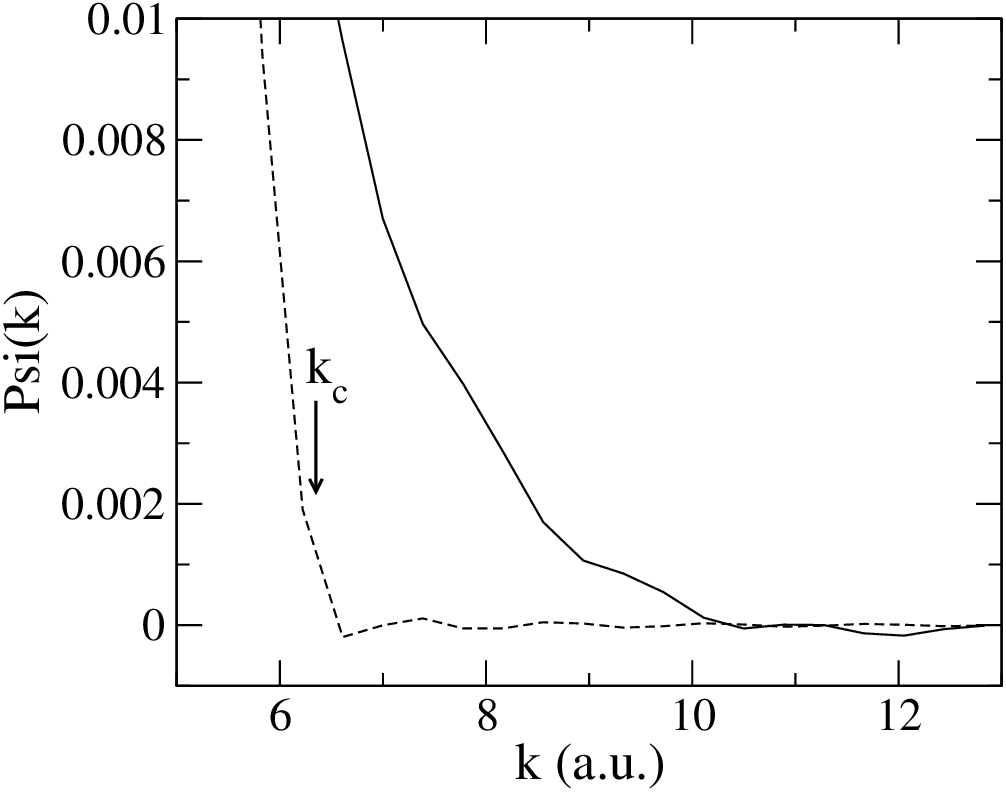}
\caption{
   Filtered (dashed lines) and unfiltered (full lines) oxygen 2$p$ pseudo
atomic orbital, generated with a strict cutoff $r_c=3.94$ bohr as 
proposed by Sankey and Niklewski~\cite{Sankey-Niklewski1989}.
   It was filtered with a cutoff $\kappa=25$, which corresponds 
to a plane wave cutoff $k_c=6.35$ bohr$^{-1}$, or $k_c^2=40$ Ryd.
   Upper panel: real space shape.
   Lower panel: tails of their Fourier transform.
}
\label{Op1}
\end{figure}

   Fig.~\ref{Op1} shows, as an example, the unfiltered and filtered
oxygen 2$p$ pseudo atomic orbital, generated as proposed by Sankey
and Niklewski~\cite{Sankey-Niklewski1989,Soler2002} with a
Troullier-Martins pseudopotential~\cite{Troullier-Martins1991}.
   To enhance the filtering effect, we have used a very small filter
cutoff.
   Still, it may be seen that the Fourier components above the cutoff 
are very efficiently suppressed, although this is achieved (with this 
small cutoff) at the expense of a substantial change in its shape.

   Finally, the most confined eigenfunctions, $\phi_1(x)$, for
each angular momentum $l$, may be used to generate a localized
distribution with given multipole moments, as required in the
ultrasoft pseudopotential~\cite{Vanderbilt1990} and projector 
augmented waves~\cite{Blochl1994} methods, among others problems 
in computational physics~\cite{Hockney-Eastwood1988}.
   They may be used also as a basis of localized orbitals, for the 
expansion of the electron wavefunctions~\cite{Gan-Haynes-Payne2001},
which is asymptotically complete, within the confining spheres,
as the filter cutoff increases.

\section{Application within SIESTA}

   There are three contributions to the eggbox effect in SIESTA
(an artificial rippling of the total energy surface as a function
of the positions of the atoms relative to the integration grid points):
   {\it i)} the so-called neutral-atom potential~\cite{Soler2002}
$V_{NA}(\r)$ given by the local part of the atomic pseudopotentials 
minus the Hartree potential of the free-atom electron densities;
   {\it ii)} the exchange and correlation potential $V_{xc}(\r)$, 
given by the electron valence density $\rho(\r)$, which in turn 
is given by a sum of products of atomic basis orbitals 
$\varphi_\mu(\r)$.
   These two contributions are frequently comparable in magnitude;
   {\it iii)} the nonlocal core correction (NLCC) to $V_{xc}(\r)$,
given by a pseudocore electron density $\rho_{NLCC}(\r)$ added
to $\rho(\r)$. 
   This added density is generally very large and localized and, 
when the NLCC is present, it normally dominates the eggbox effect.
   Finally, the Hartree energy, given by the self-interaction of
$\rho(\r)$, also contributes to the eggbox but, 
since the Hartree potential is much smoother than the density,
this contribution is always negligible compared to the other ones.

   Thus, in order to cut drastically the eggbox effect, we must
filter $\rho_{NLCC}(\r)$, $V_{NA}(\r)$, and $\varphi_\mu(\r)$.
   The first two may be filtered with the plane wave cutoff $k_c$
of the real-space integration grid used to calculate the matrix
elements of $V_{NA}(\r)$ and $V_{xc}(\r)$.
   The filtering cutoff required for $\varphi_\mu(\r)$ is somewhat
less clear, because we need to treat products of two $\varphi$'s in
the integration grid, not just the $\varphi$'s themselves.
   In principle, the plane wave cutoff of a product of two 
functions is twice that of the functions themselves, what would
suggest that $\varphi_\mu(\r)$ should be filtered with $k_c/2$.
   However, a widespread experience with plane wave codes has 
shown that this criterion is too strict, and that in practice
the effective cutoff for the density is typically less than two
times that of the wavefunctions.
   Therefore, we have checked that making the filter cutoff for
$\varphi_\mu(\r)$ equal to $\sim 0.6 k_c$ leads generally to
the best convergence, as a function of $k_c$.

\begin{figure}[htpb]
\includegraphics[width=0.9\columnwidth,clip]{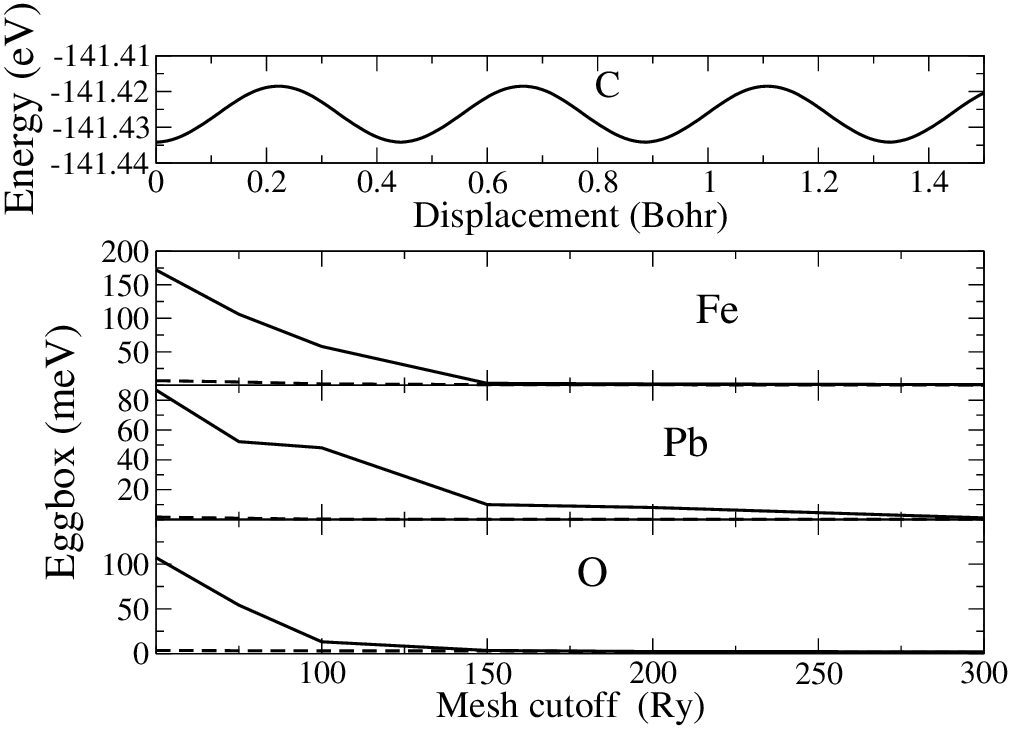}
\caption{
   a) Total energy of an isolated carbon atom, as it is displaced
in a large unit cell, using an integration grid with a plane wave 
cutoff $k_c^2 = 50$ Ryd, whose points are separated by 
$\Delta x = \pi/k_c = 0.44$ bohr.
   b) Magnitude of the eggbox effect (peak to peak of total energy)
for several isolated atoms with hard pseudopotentials or nonlocal
core corrections (in Pb and Fe), as a function of the plane wave
cutoff of the integration mesh.
   Full lines: without filtering. 
   Dashed lines: with filtering.
}
\label{eggbox}
\end{figure}

\begin{figure}[htpb]
\includegraphics[width=0.9\columnwidth,clip]{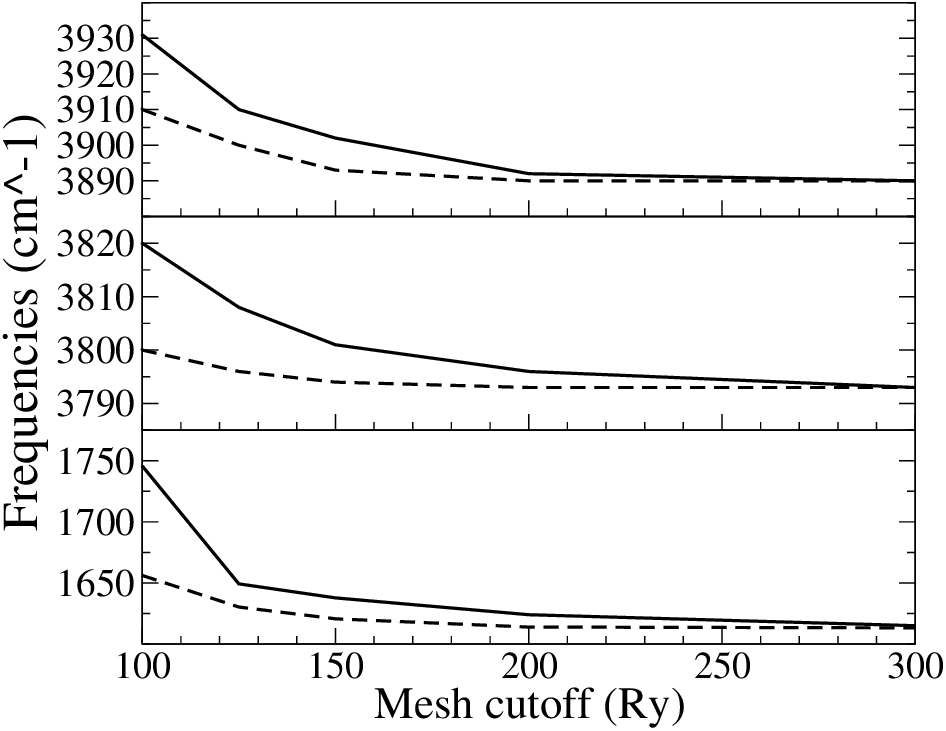}
\caption{
   Vibrational frequencies of the water molecule, calculated from
the hessian matrix, which was obtained by finite differences from 
the forces on the atoms displaced from their equilibrium positions.
   The $x$ axis is the plane wave cutoff of the integration grid
used in SIESTA.
   Full lines: without filtering. 
   Dashed lines: with filtering.   
}
\label{h2o}
\end{figure}

   Figure~\ref{eggbox} shows the eggbox effect of isolated atoms 
displaced across the integration mesh.
   It may be seen that the effect is indeed eliminated almost 
completely by filtering.
   Of course, we shall not eliminate the eggbox effect at the expense
of filtering the pseudopotentials and basis functions so much as to
change the physical results.
   Figure~\ref{h2o} shows the vibrational frequencies of the water
molecule, calculated by diagonalizing the dynamical matrix obtained
by finite differences~\cite{Paulsson2005}.
   As the plane wave cutoff $k_c$ of the integration grid is reduced,
$\rho_{NLCC}(\r)$, $V_{NA}(\r)$ are filtered with that cutoff, and
$\varphi_\mu(\r)$ is filtered with $0.7 k_c$.
   It may be seen that much lower cutoffs are required, to converge
accurate frequencies, with than without filtering.

\section{Conclusions}

   We have presented a general method to generate distributions,
with a given angular momentum, which are optimally confined within
a strict cutoff in both real and Fourier space.
   They can be used by themselves, as to produce localized
distributions with given multipole moments, or as a basis for
expanding and filtering an arbitrary initial distribution.
   As an example, we have shown how they can be used to filter
the pseudopotentials and basis functions in the density functional 
method SIESTA, thus eliminating the eggbox effect on the total energy, 
due to the calculation of matrix elements in a real space integration 
grid.

\begin{acknowledgments}
   We want to thank Alberto Garc\'{i}a for useful discussions and
M. Fern\'{a}ndez-Serra for the basis set of the water molecule 
\cite{FernandezSerra2004}.
   This work has been founded by grant BFM2003-03372 from the
Spanish Ministery of Science.
\end{acknowledgments}

\appendix

\section{Variational principles}

   Here we show that the filtering basis functions $\phi_i(r)$
obey a simple variational principle, and we also present an
alternative principle for gaussian basis functions.
   It may be easily shown, by a straightforward functional
derivative, that the eigenvalue equation (\ref{eigsystem2})
is equivalent to the variational principle
\begin{equation}
\int_0^1 \int_0^1 dx ~dx' \phi(x) K^2(x,x') \phi(x') = \mbox{max}
\label{K2phiphi}
\end{equation}
subject to the condition of normalization of $\phi(x)$ 
within $0 \le x \le 1$.
   Now, using that the Fourier transform of $\phi(x)$ is
\begin{equation}
g(y) = \int_0^1 dx ~\phi(x) ~K(x,y)
\label{gofy}
\end{equation}
as well as the definition (\ref{K2ofK}) and the fact that the
total norm of a function is the same in real and Fourier space:
\begin{eqnarray}
\lefteqn{ \int_0^1 dy \int_0^1 dx \phi(x) K(x,y)
                  \int_0^1 ~dx' K(y,x') \phi(x') } \nonumber \\
&=& \int_0^1 dy ~g^2(y)
 =  1 - \int_1^\infty dy ~g^2(y) = \mbox{max} \nonumber \\
&\Rightarrow& \int_1^\infty dy ~g^2(y) = \mbox{min}
\label{var1}
\end{eqnarray}
   Thus, our basis functions $\phi(x)$ are the normalized 
distributions which are strictly confined to $0 \le x \le 1$ 
(i.\ e.\ $r \le r_c$) and whose Fourier transform has the smallest 
norm in $y>1$ ($k > k_c$).

   Interestingly, an alternative variational principle may be
demonstrated for a basis of gaussian functions.
   Thus, we maximize the
confinement of a normalized distribution $\phi(\r)$ and its Fourier
transform $\phi(\k)$, in the sense of least squared dispersion: 
\begin{equation}
 \frac{1}{r_c^2} \int d\r ~\r^2 \phi^2(\r) + 
 \frac{1}{k_c^2} \int d\k ~\k^2 \phi^2(\k) 
     = \mbox{min}
\label{var2}
\end{equation}
where $r_c$ and $k_c$ are here scale factors that determine the relative
confinement in real and Fourier space, rather than strict cutoffs.
   Multiplying by $k_c^2/2$:
\begin{equation}
 \int d\r \frac{k_c^2 \r^2}{2 r_c^2} \phi^2(\r) + 
 \int d\k \frac{\k^2}{2} \phi^2(\k) 
     = \mbox{min}
\label{var3}
\end{equation}
   Now, the first term is the potential energy of a quantum 
harmonic oscillator with spring constant $(k_c/r_c)^2$ and wave function
$\phi(\r)$, and the second term is its kinetic energy.
   Its well known solutions are gaussians times Hermite polynomials
\cite{Cohen1977}.
   A similar (but not orthonormal) basis, made of gaussians times powers 
of $r$, was used by Hartwigsen {\it et al}
~\cite{Hartwigsen-Goedecker-Hutter1998}
to generate compact separable pseudopotentials.

\bibliographystyle{apsrev}
\bibliography{dft,siesta,misc,filter}

\end{document}